\documentclass[11pt]{article}
\usepackage{eurosym}
\usepackage{amssymb,amsmath,amsfonts}
\usepackage{graphicx}
\usepackage{graphics}
\usepackage{eepic,epsfig}


\begin{document}

\title{Electromagnetic vacuum densities induced by a cosmic string }
\author{A. A. Saharian$^{1}$,\thinspace\ V. F. Manukyan$^{2}$,\thinspace\ N.
A. Saharyan$^{1}$ \\
\\
\textit{$^1$Department of Physics, Yerevan State University,}\\
\textit{1 Alex Manoogian Street, 0025 Yerevan, Armenia}\vspace{0.3cm}\\
\textit{$^{2}$Faculty of Natural Science and Mathematics, Shirak State
University,}\\
\textit{4 Paruyr Sevak Street, 3126 Gyumri, Armenia}}
\maketitle

\begin{abstract}
We investigate the influence of a generalized cosmic string in $(D+1)$%
-dimensional spacetime on the local characteristics of the electromagnetic
vacuum. Two special cases are considered with flat and locally de Sitter
background geometries. The topological contributions in the vacuum
expectation values (VEVs) of the squared electric and magnetic fields are
explicitly separated. Depending on the number of spatial dimensions and on
the planar angle deficit induced by the cosmic string, these contributions
can be either negative or positive. In the case of the flat bulk, the
VEV of the energy-momentum tensor is evaluated as well. For the locally de Sitter
bulk, the influence of the background gravitational field essentially
changes the behavior of the vacuum densities at distances from the string
larger than the curvature radius of the spacetime.
\end{abstract}

\bigskip

\section{Introduction}

The formation of topological defects is one of the interesting consequences
of symmetry breaking phase transitions in the early Universe. Depending on
the topology of the vacuum manifold these are domains walls, strings,
monopoles and textures. Among them, cosmic strings have been of increasing
interest due to the importance that they may have in cosmology \cite{Vile94}%
. This class of defects are sources of a number of interesting astrophysical
effects such as the generation of gravitational waves, high-energy cosmic
rays, and gamma ray bursts. Another interesting effect is the influence of
cosmic strings on the temperature anisotropies of the cosmic microwave
background radiation. More recently, a mechanism for the generation of the
cosmic string type objects is proposed within the framework of brane
inflation \cite{Hind11,Cope11}.

Depending on the underlying microscopic model, the cosmic strings can be
either nontrivial field configurations or more fundamental objects in
superstring theories. In the simplest theoretical model describing the
straight cosmic strings the influence of the latter on the surrounding
geometry at large distances from the core is reduced to the generation of a
planar angle deficit. In quantum field theory, among the most interesting
effects of the corresponding nontrivial spatial topology is the vacuum
polarization. This effects have been discussed for scalar, fermion and
vector fields (see, for instance, references cited in \cite{Bell14,Mota17}).

In the present paper we present the results of the investigations for the
influence of a cosmic string on the vacuum electromagnetic fluctuations. The
vacuum expectation value (VEV) of the energy-momentum tensor for the
electromagnetic field around a cosmic string in $D=3$ spatial dimensions has
been obtained in \cite{Frol87,Dowk87} on the base of the Green function. For
superconducting cosmic strings, assuming that the string is surrounded by a
superconducting cylindrical surface, in \cite{Brev95} the electromagnetic
energy produced in the lowest mode is evaluated. The VEVs of the squared
electric and magnetic fields, and of the energy-momentum tensor for the
electromagnetic field inside and outside of a conducting cylindrical shell
in the cosmic string spacetime have been investigated in \cite{Beze07}. The
corresponding VEVs, the Casimir-Polder and the Casimir forces in the
geometry of two parallel conducting plates on the background of cosmic
string spacetime were discussed in \cite{Beze12,Beze12c}. The repulsive
Casimir-Polder forces acting on a polarizable microparticle in the geometry
of a straight cosmic string are investigated in \cite{Bard10,Saha11}. The
Casimir-Polder interaction between an atom and a metallic cylindrical shell
in cosmic string spacetime has been studied in \cite{Saha12} (see also \cite%
{Saha12b}). The electromagnetic field correlators and the VEVs of the
squared electric and magnetic fields around a cosmic string in background of
$(D+1)$-dimensional locally de Sitter (dS) spacetime were evaluated in \cite%
{Saha17} (for quantum vacuum effects in the geometry of two straight
parallel cosmic strings see \cite{Muno14} and references therein).

The organization of the paper is as follows. In the next section we present
the complete set of the electromagnetic field mode functions on the bulk of
a $(D+1)$-dimensional generalized cosmic string geometry. In Section \ref%
{sec:E2}, by using these mode functions, the VEVs of the squared electric
and magnetic fields are investigated. The VEV of the energy-momentum tensor
is studied in Section \ref{sec:EMT}. In Section \ref{sec:dS} we consider the
VEVs of the squared electric and magnetic fields, and of the vacuum energy
density for a cosmic string in locally dS spacetime. The main results are
summarized in Section \ref{sec:Conc}.

\section{Electromagnetic field modes around a cosmic string in flat spacetime%
}

\label{sec:Mink}

In the first part of the paper we consider a quantum electromagnetic field
with the vector potential $A_{\mu }(x)$ in the background of a $(D+1)$%
-dimensional flat spacetime, in the presence of an infinitely long straight
cosmic string with the line element
\begin{equation}
ds^{2}=dt^{2}-dr^{2}-r^{2}d\phi ^{2}-\left( d\mathbf{z}\right) ^{2},
\label{ds2M}
\end{equation}%
where $\mathbf{z=}\left( z^{3},...,z^{D}\right) ,$ $0\leq \phi \leq \phi
_{0} $ and the points $(r,\phi ,\mathbf{z})$ and $(r,\phi +\phi _{0},\mathbf{%
z})$ are to be identified. The geometry described by (\ref{ds2M}) is flat
everywhere except the points on the axis $r=0$ where one has a delta-type
singularity. Though the local characteristics of the cosmic string spacetime
in the region $r>0$ are the same as those in flat spacetime, for $\phi
_{0}\neq 2\pi $ these manifolds differ globally. We are interested in the
influence of nontrivial topology induced by a planar angle deficit on local
characteristics of the electromagnetic vacuum. In the canonical quantization
procedure we need to know a complete orthonormal set $\{A_{(\beta )\mu
},A_{(\beta )\mu }^{\ast }\}$ of solutions to the classical field equations $%
\partial _{\nu }\left( \sqrt{|g|}F^{\mu \nu }\right) =0$, where $F_{\mu \nu
} $ is the electromagnetic field tensor, $F_{\mu \nu }=\partial _{\mu
}A_{\nu }-\partial _{\nu }A_{\mu }$, and $g$ is the determinant of the
metric tensor. Here and below the set $(\beta )$ of quantum numbers
specifies the mode functions. In the Coulomb gauge one has $A_{(\beta )0}=0$%
, $\partial _{l}(\sqrt{|g|}A_{(\beta )}^{l})=0$, $l=1,...,D$.

In the problem at hand the set of quantum numbers is specified as $(\beta
)=(\gamma ,m,\mathbf{k},\sigma )$. Here $0\leqslant \gamma <\infty $ is the
radial quantum number, $m=0,\pm 1,\pm 2,\ldots $ is the azimuthal quantum
number, $\mathbf{k}=(k_{3},\ldots ,k_{D})$ is the momentum in the subspace $%
\left( z^{3},...,z^{D}\right) $, and $\sigma =1,\ldots ,D-1$ enumerates the
polarization states. The cylindrical modes for the electromagnetic field are
given by%
\begin{eqnarray}
A_{(\beta )\mu } &=&C_{(\beta )}\left( 0,\frac{iqm}{r},-r\partial
_{r},0,\ldots ,0\right) J_{q|m|}(\gamma r)e^{i(qm\phi +\mathbf{k}\cdot
\mathbf{z}-\omega t)},\;\sigma =1,  \notag \\
A_{(\beta )\mu } &=&C_{(\beta )}\omega \left( 0,\epsilon _{\sigma l}+i\frac{%
\mathbf{k}\cdot \boldsymbol{\epsilon }_{\sigma }}{\omega ^{2}}\partial
_{l}\right) J_{q|m|}(\gamma r)e^{i(qm\phi +\mathbf{k}\cdot \mathbf{z}-\omega
t)},\;\sigma =2,\ldots ,D-1,  \label{A1M}
\end{eqnarray}%
where $l=1,\ldots ,D$, $q=2\pi /\phi _{0}$, $\omega =\sqrt{\gamma ^{2}+k^{2}}
$, $k^{2}=\sum_{l=3}^{D}k_{l}^{2}$, and $J_{\nu }(x)$ is the Bessel function
of the first kind. The scalar products are given as $\mathbf{k}\cdot \mathbf{%
z}=\sum_{l=3}^{D}k_{l}z^{l}$ and $\mathbf{k}\cdot \boldsymbol{\epsilon }%
_{\sigma }=\sum_{l=3}^{D}k_{l}\epsilon _{\sigma l}$. For the polarization
vector $\boldsymbol{\epsilon }_{\sigma }$ one has $\epsilon _{\sigma
1}=\epsilon _{\sigma 2}=0$, $\sigma =2,\ldots ,D-1$, and the relations%
\begin{eqnarray}
\sum_{l,n=3}^{D}\left( \omega ^{2}\delta _{nl}-k_{l}k_{n}\right) \epsilon
_{\sigma l}\epsilon _{\sigma ^{\prime }n} &=&\gamma ^{2}\delta _{\sigma
\sigma ^{\prime }},  \notag \\
\omega ^{2}\sum_{\sigma =2}^{D-1}\epsilon _{\sigma n}\epsilon _{\sigma
l}-k_{n}k_{l} &=&\gamma ^{2}\delta _{nl},\;l,n=3,...,D.  \label{Pol}
\end{eqnarray}%
The polarization state $\sigma =1$ is the mode of TE type and $\sigma
=2,\ldots ,D-1$ correspond to $D-2$ modes of the TM\ type.

The coefficients $C_{(\beta )}$ are determined by the normalization
condition for vector fields. For a general background with the metric tensor
$g_{ik}$ this condition is written as%
\begin{equation}
\int d^{D}x\sqrt{|g|}g^{00}[A_{(\beta ^{\prime })\nu }^{\ast }(x)\nabla
_{0}A_{(\beta )}^{\nu }(x)-(\nabla _{0}A_{(\beta ^{\prime })\nu }^{\ast
}(x))A_{(\beta )}^{\nu }(x)]=4i\pi \delta _{\beta \beta ^{\prime }},
\label{NormCond}
\end{equation}%
where $\nabla _{\mu }$ stands for the covariant derivative and $\delta
_{\beta \beta ^{\prime }}$ is understood as the Kronecker symbol for
discrete components of the collective index $\beta $ and the Dirac delta
function for the continuous ones. In the problem under consideration, by
using the standard integral for the product of Bessel functions, one finds
\begin{equation}
|C_{(\beta )}|^{2}=\frac{q}{(2\pi )^{D-2}\gamma \omega },  \label{Normc}
\end{equation}%
for all the polarizations $\sigma =1,\ldots ,D-1$.

With a given set of mode functions (\ref{A1M}), the VEV of any physical
quantity $F\{A_{\mu }(x),A_{\nu }(x)\}$ bilinear in the field is evaluated
by making use of the mode-sum formula%
\begin{equation}
\langle 0|F\{A_{\mu }(x),A_{\nu }(x)\}|0\rangle =\sum_{\beta }F\{A_{(\beta
)\mu }(x),A_{(\beta )\nu }^{\ast }(x)\},  \label{VEV}
\end{equation}%
where $|0\rangle $ stands for the vacuum state and%
\begin{equation}
\sum_{\beta }=\sum_{\sigma =1}^{D-1}\sum_{m=-\infty }^{\infty }\int d\mathbf{%
k}\int_{0}^{\infty }d\gamma .  \label{SumBet}
\end{equation}%
The expression in the right-hand side of (\ref{VEV}) is divergent and
requires a regularization with the subsequent renormalization. The
regularization can be done by introducing a cutoff function or by the point
splitting. A very convenient tool for studying one-loop divergences is the
heat kernel expansion (for a general introduction with applications to
conical spaces see \cite{Kirs02,Vass03}). The heat kernels of Laplacians for
higher spin fields and the related asymptotic expansions on manifolds with
conical singularities were studied in \cite{Furs97}. In what follows we will
use the appoach based on the cutoff function. Compared with the point
splitting it essentially simplifies the calculations of the topological
contributions in the VEVs of local observables.

\section{VEVs of the squared electric and magnetic fields}

\label{sec:E2}

First we consider the VEV of the squared electric field. This VEV is
obtained by making use of the mode-sum formula
\begin{equation}
\langle 0|E^{2}|0\rangle \equiv \langle E^{2}\rangle
=-g^{00}g^{il}\sum_{\beta }\partial _{0}A_{(\beta )i}(x)\partial
_{0}A_{(\beta )l}^{\ast }(x).  \label{E2mode}
\end{equation}%
Note that the VEV of the electric field squared determines the
Casimir-Polder potential between the cosmic string and a polarizable
particle with a frequency-independent polarizability. We will regularize the
VEV by introducing the cutoff function $e^{-b\omega ^{2}}$ with $b>0$ (about
using this kind of cutoff function in the evaluation of the Casimir energy
see, for example, \cite{Asor13}). Substituting the mode functions and using (%
\ref{Pol}), the regularized VEV\ is presented in the form%
\begin{eqnarray}
\langle E^{2}\rangle _{\mathrm{reg}} &=&\frac{4\left( 4\pi \right) ^{1-D/2}q%
}{\Gamma (D/2-1)}\sideset{}{'}{\sum}_{m=0}^{\infty }\int_{0}^{\infty
}dk\,k^{D-3}\int_{0}^{\infty }d\gamma \frac{\gamma }{\omega }e^{-b\omega
^{2}}  \notag \\
&&\times \left\{ \left( \gamma ^{2}+2k^{2}\right) G_{qm}(\gamma r)+\left[
(D-2)\gamma ^{2}+(D-3)k^{2}\right] J_{qm}^{2}(\gamma r)\right\} ,
\label{E2c}
\end{eqnarray}%
where the prime on the summation sign means that the term $m=0$ should be
taken with the coefficient 1/2 and the function
\begin{equation}
G_{\nu }(x)=J_{\nu }^{\prime 2}(x)+\frac{\nu ^{2}}{x^{2}}J_{\nu }^{2}(x)
\label{Gnu}
\end{equation}%
is introduced.

For the further transformation of the right-hand side in (\ref{E2c}) we use
the integral representation%
\begin{equation}
\frac{1}{\omega }=\frac{2}{\sqrt{\pi }}\int_{0}^{\infty }dy\,e^{-(\gamma
^{2}+k^{2})y^{2}}.  \label{IntRep}
\end{equation}%
After the evaluation of the $k$-integrals one gets%
\begin{eqnarray}
\langle E^{2}\rangle _{\mathrm{reg}} &=&\frac{2^{4-D}q}{\pi ^{(D-1)/2}}%
\sideset{}{'}{\sum}_{m=0}^{\infty }\int_{0}^{\infty }\frac{dy}{w^{D/2}}%
\,\int_{0}^{\infty }d\gamma \gamma \left[ \left( D-2-w\partial _{w}\right)
G_{qm}(\gamma r)\right.  \notag \\
&&\left. +\left( D-2\right) \left( \frac{D-3}{2}-w\partial _{w}\right)
J_{qm}^{2}(\gamma r)\right] e^{-w\gamma ^{2}},  \label{E21c}
\end{eqnarray}%
where $w=b+y^{2}$. Next we use the integral \cite{Prud86}
\begin{equation}
\int_{0}^{\infty }d\gamma \gamma e^{-w\gamma ^{2}}J_{qm}(\gamma
r)J_{qm}(\gamma r^{\prime })=\frac{1}{2w}\exp \left( -\frac{r^{2}+r^{\prime
2}}{4w}\right) I_{qm}\left( \frac{rr^{\prime }}{2w}\right) ,  \label{Int1}
\end{equation}%
with $r^{\prime }=r$ and with $I_{\nu }(x)$ being the modified Bessel
function. The remaining integral over $\gamma $ is written as%
\begin{equation}
\int_{0}^{\infty }d\gamma \gamma e^{-w\gamma ^{2}}G_{qm}(\gamma
r)=\lim_{r^{\prime }\rightarrow r}\left( \partial _{r}\partial _{r^{\prime
}}+\frac{q^{2}m^{2}}{rr^{\prime }}\right) \int_{0}^{\infty }d\gamma \frac{%
e^{-w\gamma ^{2}}}{\gamma }J_{qm}(\gamma r)J_{qm}(\gamma r^{\prime }).
\label{Int2}
\end{equation}%
By taking into account that $e^{-w\gamma ^{2}}=\gamma ^{2}\int_{w}^{\infty
}dt\,e^{-t\gamma ^{2}}$ and using (\ref{Int1}) one finds%
\begin{equation}
\int_{0}^{\infty }d\gamma \ \frac{e^{-w\gamma ^{2}}}{\gamma }J_{qm}(\gamma
r)J_{qm}(\gamma r^{\prime })=\frac{1}{2}\int_{0}^{1/(4w)}\frac{dx}{x}%
\,e^{-(r^{2}+r^{\prime 2})x}I_{qm}(2rr^{\prime }x).  \label{Int3}
\end{equation}%
Substituting this into (\ref{Int2}) we obtain%
\begin{equation}
\int_{0}^{\infty }d\gamma \gamma e^{-w\gamma ^{2}}G_{qm}(\gamma r)=\frac{u}{%
r^{2}}\left( \partial _{u}+1\right) e^{-u}I_{qm}(u),  \label{Int4}
\end{equation}%
with the notation $u=r^{2}/(2w)$.

By taking into account (\ref{Int1}), (\ref{Int4}) and passing in (\ref{E21c}%
) from the integration over $y$ to the integration over $u$ one finds%
\begin{eqnarray}
\langle E^{2}\rangle _{\mathrm{reg}} &=&\frac{4q}{\left( 2\pi \right)
^{(D-1)/2}r^{D+1}}\int_{0}^{r^{2}/(2b)}du\,\frac{u^{(D-1)/2}}{\sqrt{%
1-2bu/r^{2}}}\left[ \left( D-2+\partial _{u}u\right) \left( \partial
_{u}+1\right) \right.  \notag \\
&&\left. +\left( D-2\right) \left( \frac{D-3}{2}+\partial _{u}u\right) %
\right] e^{-u}\sideset{}{'}{\sum}_{m=0}^{\infty }I_{qm}\left( u\right) .
\label{E22c}
\end{eqnarray}%
For the further transformation we use the formula \cite{Beze12,Beze12b}%
\begin{equation}
\sideset{}{'}{\sum}_{m=0}^{\infty }I_{qm}\left( u\right) =\frac{1}{q}%
\sideset{}{'}{\sum}_{l=0}^{[q/2]}e^{u\cos (2l\pi /q)}-\frac{1}{2\pi }%
\int_{0}^{\infty }dy\frac{\sin (q\pi )e^{-u\cosh y}}{\cosh (qy)-\cos (q\pi )}%
,  \label{Summ}
\end{equation}%
where $[q/2]$ stands for the integer part of $q/2$ and the prime on the
summation sign means that the terms $l=0$ and $l=q/2$ (for even values of $q$%
) should be taken with additional coefficient 1/2. In the case $q=1$ the $%
l=0 $ term remains only. From here it follows that the contribution of the
term $l=0$ in the VEV (\ref{E22c}) corresponds to the VEV in Minkowski
spacetime in the absence of cosmic string. We will denote the corresponding
regularized VEV by $\langle E^{2}\rangle _{\mathrm{reg}}^{(0)}$. The latter
is obtained from (\ref{E22c}) taking the $l=0$ term in (\ref{Summ}) instead
of the series over $m$:%
\begin{equation}
\langle E^{2}\rangle _{\mathrm{reg}}^{(0)}=\frac{\left( D-1\right) \Gamma
((D+1)/2)}{2^{D-1}\pi ^{D/2-1}\Gamma (D/2)b^{(D+1)/2}}.  \label{Ereg0}
\end{equation}

The remaining part in (\ref{E22c}) is the contribution induced by the cosmic
string (topological part). This part is finite in the limit $b\rightarrow 0$
and the cutoff can be removed. We denote the topological part in the VEV of
the squared electric field as $\langle E^{2}\rangle _{\mathrm{t}}$:%
\begin{equation}
\langle E^{2}\rangle _{\mathrm{t}}=\lim_{b\rightarrow 0}\left[ \langle
E^{2}\rangle _{\mathrm{reg}}-\langle E^{2}\rangle _{\mathrm{reg}}^{(0)}%
\right] .  \label{E2t}
\end{equation}%
Substituting (\ref{Summ}) into (\ref{E22c}), separating the $l=0$ term and
taking the limit $b\rightarrow 0$, after the evaluation of the $u$-integral,
for the topological contribution we find the following result%
\begin{equation}
\langle E^{2}\rangle _{\mathrm{t}}=-\frac{\Gamma (\left( D+1\right) /2)}{%
\left( 4\pi \right) ^{(D-1)/2}r^{D+1}}\left[ \left( D-1\right)
c_{D+1}+2\left( D-3\right) c_{D-1}\right] ,  \label{E23}
\end{equation}%
with the notation%
\begin{equation}
c_{n}(q)=\sideset{}{'}{\sum}_{l=1}^{[q/2]}\frac{1}{\sin ^{n}(\pi l/q)}-\frac{%
q}{\pi }\sin (q\pi )\int_{0}^{\infty }dy\frac{\cosh ^{-n}y}{\cosh (2qy)-\cos
(q\pi )},  \label{cnqu}
\end{equation}%
where the prime means that for even values of $q$ the term $l=[q/2]$ should
be taken with coefficient 1/2. Note that the functions (\ref{cnqu}) also
appear in the coefficients of the heat kernel expansion in the background of
cosmic string spacetime (see \cite{Dowk87b,Furs94}).

\begin{figure}[tbph]
\begin{center}
\epsfig{figure=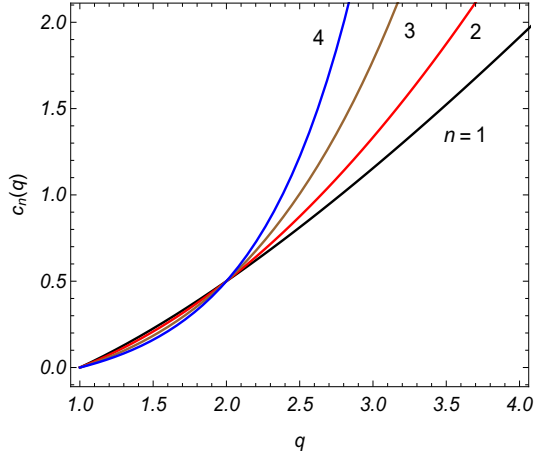,width=7.cm,height=6.cm}
\end{center}
\caption{The functions $c_{n}(q)$ for different values of $n$ (the numbers
near the curves).}
\label{fig1}
\end{figure}

In figure \ref{fig1} we have plotted the functions $c_{n}(q)$ for different
values of $n$. These functions are monotonically increasing positive
functions of $q>1$. For them one has $c_{n}(2)=1/2$ and $c_{n}(q)>c_{n+1}(q)$
for $1<q<2$. In the region $q>2$ we have $c_{n}(q)<c_{n+1}(q)$.

From (\ref{E23}) we conclude that the topological part in the VEV of the
squared electric field is negative for $D\geqslant 3$ and for $q=2$ one gets%
\begin{equation}
\langle E^{2}\rangle _{\mathrm{t}}=-\frac{\Gamma (\left( D+1\right) /2)}{%
2\left( 4\pi \right) ^{(D-1)/2}r^{D+1}}\left( 3D-7\right) ,\;q=2.
\label{E2q2}
\end{equation}

For even values of $n$, the function $c_{n}(q)$ can be further simplified by
using the recurrence scheme described in \cite{Beze06}. In particular, one
has $c_{2}(q)=(q^{2}-1)/6$ and
\begin{eqnarray}
c_{4}(q) &=&\frac{q^{2}-1}{90}\left( q^{2}+11\right) ,  \notag \\
c_{6}(q) &=&\frac{q^{2}-1}{1890}(2q^{4}+23q^{2}+191).  \label{c6q}
\end{eqnarray}%
By using these results one gets \cite{Beze07}%
\begin{equation}
\langle E^{2}\rangle _{\mathrm{t}}=-\frac{q^{2}-1}{180\pi r^{4}}\left(
q^{2}+11\right) ,  \label{E2D3}
\end{equation}%
for $D=3$ and
\begin{equation}
\langle E^{2}\rangle _{\mathrm{t}}=-\frac{q^{2}-1}{1890\pi ^{2}r^{6}}\left(
q^{4}+22q^{2}+211\right) ,  \label{E2D5}
\end{equation}%
for $D=5$. The topological part induced by the string is negative for $q>1$.

Now we consider the VEV of the squared magnetic field, given by
\begin{equation}
\langle B^{2}\rangle =\frac{1}{2}g^{lm}g^{np}\langle F_{ln}F_{mp}\rangle =%
\frac{1}{2}g^{lm}g^{np}\sum_{\beta }F_{\left( \beta \right) ln}F_{\left(
\beta \right) mp}^{\ast },  \label{B2}
\end{equation}%
where the summation goes over the spatial indices and $F_{\left( \beta
\right) ln}=\partial _{l}A_{(\beta )n}-\partial _{n}A_{(\beta )l}$. Note
that for $D>3$ the magentic field is not a spatial vector. With the mode
functions from (\ref{A1M}), the VEV, regularized by the cutoff function $%
e^{-b\omega ^{2}}$, $b>0$,\ is presented as%
\begin{eqnarray}
\langle B^{2}\rangle _{\mathrm{reg}} &=&\frac{4\left( 4\pi \right) ^{1-D/2}q%
}{\Gamma (D/2-1)}\sideset{}{'}{\sum}_{m=0}^{\infty }\int_{0}^{\infty
}dk\,k^{D-3}\int_{0}^{\infty }d\gamma \frac{\gamma }{\omega }e^{-b\omega
^{2}}  \notag \\
&&\times \left\{ \left[ \left( D-2\right) \gamma ^{2}+2k^{2}\right]
G_{qm}(\gamma r)+\left[ \gamma ^{2}+\left( D-3\right) k^{2}\right]
J_{qm}^{2}(\gamma r)\right\} .  \label{B21}
\end{eqnarray}%
The further evaluation is similar to that for the squared electric field and
we will omit the details. The final result for the topological contribution
\begin{equation}
\langle B^{2}\rangle _{\mathrm{t}}=\lim_{b\rightarrow 0}\left[ \langle
B^{2}\rangle _{\mathrm{reg}}-\langle B^{2}\rangle _{\mathrm{reg}}^{(0)}%
\right] ,  \label{B2t}
\end{equation}%
is given by the expression%
\begin{equation}
\langle B^{2}\rangle _{\mathrm{t}}=\frac{2\Gamma (\left( D+1\right) /2)}{%
\left( 4\pi \right) ^{(D-1)/2}r^{D+1}}\left[ \left( D-3\right) \left(
D-2\right) c_{D-1}(q)-\frac{D-1}{2}c_{D+1}(q)\right] ,  \label{B22}
\end{equation}%
with the function $c_{n}(q)$ from (\ref{cnqu}). For the regularized VEV\ in
the absence of cosmic string one has $\langle B^{2}\rangle _{\mathrm{reg}%
}^{(0)}=\langle E^{2}\rangle _{\mathrm{reg}}^{(0)}$. For the special case $%
q=2$, by taking into account that $c_{n}(2)=1/2$, we find%
\begin{equation}
\langle B^{2}\rangle _{\mathrm{t}}=\frac{\Gamma (\left( D+1\right) /2)}{%
2\left( 4\pi \right) ^{(D-1)/2}r^{D+1}}\left( 2D^{2}-11D+13\right) ,\;q=2.
\label{B2q2}
\end{equation}%
Depending on $q$ and $D$ the VEV $\langle B^{2}\rangle _{\mathrm{t}}$ can be
either positive or negative. In particular, for $q=2$ one has $\langle
B^{2}\rangle _{\mathrm{t}}<0$ for $D=3$ and $\langle B^{2}\rangle _{\mathrm{t%
}}>0$ for $D\geqslant 4$.

\begin{figure}[tbph]
\begin{center}
\epsfig{figure=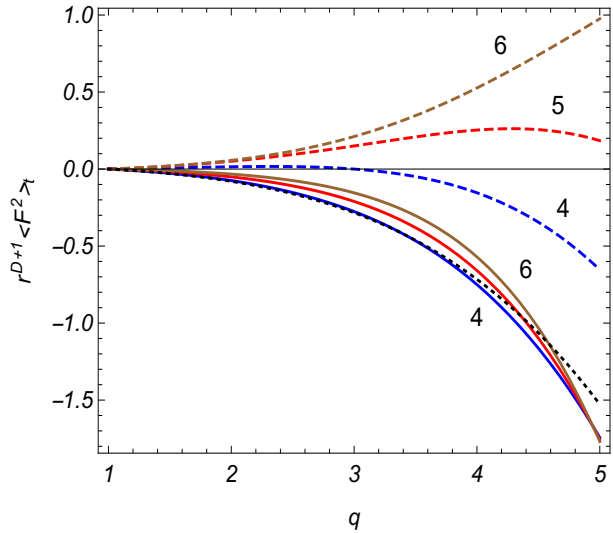,width=8.cm,height=7.cm}
\end{center}
\caption{Topological contributions in the VEVs of the squared electric and
magnetic fields, multiplied by $r^{D+1}$, as functions of $q$ for separate
values of the spatial dimensions $D=4,5,6$ (the numbers near the curves).
The full/dashed curves correspond to the electric/magnetic fields. The
dotted line presents the VEVs for $D=3$.}
\label{fig2}
\end{figure}

Simple expressions are obtained for odd values of the spatial dimension $D$.
In particular, for $D=3,5$ one gets%
\begin{eqnarray}
\langle B^{2}\rangle _{\mathrm{t}} &=&-\frac{q^{2}-1}{180\pi r^{4}}\left(
q^{2}+11\right) ,\;D=3,  \notag \\
\langle B^{2}\rangle _{\mathrm{t}} &=&-\frac{q^{2}-1}{1890\pi ^{2}r^{6}}%
\left( q^{4}-20q^{2}-251\right) ,\;D=5.  \label{B2D5}
\end{eqnarray}%
In $D=3$ the VEVs of the squared electric and magnetic fields coincide \cite%
{Beze07}. In figure \ref{fig2} we have displayed the topological
contributions in the VEVs of the squared electric and magnetic fields,
multiplied by $r^{D+1}$, as functions of the parameter $q$ for different
values of the spatial dimension $D$ (the numbers near the curves). The full
and dashed curves correspond to the electric and magnetic fields,
respectively, and the dotted curve presents the VEVs for $D=3$.

Having the VEVs of the squared electric and magnetic fields we can evaluate
the VEV of the Lagrangian density%
\begin{equation}
\langle L\rangle =-\frac{1}{16\pi }g^{\mu \rho }g^{\nu \sigma }\langle
F_{\mu \nu }F_{\rho \sigma }\rangle =\frac{\langle E^{2}\rangle -\langle
B^{2}\rangle }{8\pi }.  \label{L}
\end{equation}%
The quantity $g^{\mu \rho }g^{\nu \sigma }\langle F_{\mu \nu }F_{\rho \sigma
}\rangle $ is the Abelian analog of the gluon condensate in quantum
chromodynamics (see, for instance, \cite{Ioff10}). Note that in a number of
models (for example, string effective gravity \cite{Damo94,Saha99} and
models for generation of primordial magnetic fields during inflation \cite%
{Durr13}) the Lagrangian density contains terms of the form $f(\Phi )F_{\mu
\nu }F^{\mu \nu }$ that couple the gauge field to a scalar field $\Phi $
(dilaton field in string effective gravity and inflaton in models of
magnetic field generation). In these models, the appearance of the nonzero
VEV $\langle F_{\mu \nu }F^{\mu \nu }\rangle $ induces a contribution to the
effective potential for the field $\Phi $. The stabilization of the dilaton
field during the cosmological expansion, based on the nontrivial coupling of
dilaton to other fields, has been discussed in \cite{Damo94}. By using (\ref%
{E23}) and (\ref{B22}), for the topological contribution in the VEV of the
Lagrangian density one gets%
\begin{equation}
\langle L\rangle _{\mathrm{t}}=-\frac{\left( D-3\right) \Gamma (\left(
D+1\right) /2)}{\left( 4\pi \right) ^{(D+1)/2}r^{D+1}}\left( D-1\right)
c_{D-1}(q).  \label{L1}
\end{equation}%
This VEV vanishes for $D=3$ and is negative for $D\geqslant 4$.

The modification of the vacuum fluctuations for the electromagnetic field by
a cosmic string gives rise to the Casimir-Polder forces acting on a neutral
polarizable microparticle placed close to the string. For the general case
of anisotropic polarizability the corresponding potential in the geometry of
a $D=3$ cosmic string has been derived in \cite{Saha11}. For a special case
of isotropic polarizability $\alpha _{P}\left( \omega \right) $, the
corresponding formula takes the form%
\begin{equation}
U(r)=\frac{r^{-4}}{8\pi }\left[ \sideset{}{'}{\sum}_{l=1}^{[q/2]}\frac{%
h(r,s_{l})}{s_{l}^{4}}-\frac{q}{\pi }\sin (q\pi )\int_{0}^{\infty }dy\frac{%
h(r,\cosh y)\cosh ^{-4}y}{\cosh (2qy)-\cos (q\pi )}\right] ,  \label{CPU}
\end{equation}%
with the function%
\begin{equation}
h(r,y)=\int_{0}^{\infty }dx\,\,e^{-x}\alpha _{P}(ix/(2ry))\left[ y^{2}\left(
1+x-x^{2}\right) +x^{2}\right] .  \label{hry}
\end{equation}%
If the dispersion of the polarizability can be neglected, one gets $%
h(r,y)=2\alpha _{P}$, with $\alpha _{P}$ being the static polarizability,
and consequently%
\begin{equation}
U(r)=\alpha _{P}\frac{q^{2}-1}{360\pi r^{4}}\left( q^{2}+11\right) .
\label{CPUst}
\end{equation}%
For $\alpha _{P}>0$ the corresponding force is repulsive.

\section{Energy-momentum tensor}

\label{sec:EMT}

Another important local characteristic of the vacuum state is the VEV of the
energy-momentum tensor. It determines the back-reaction of quantum effects
on the background geometry. The VEV is evaluated by using the formula%
\begin{equation}
\langle T_{\mu }^{\nu }\rangle =-\frac{1}{4\pi }C_{\mu }^{\nu }-\delta _{\mu
}^{\nu }\langle L\rangle ,  \label{T}
\end{equation}%
where%
\begin{equation}
C_{\mu }^{\nu }=g^{\nu \kappa }g^{\rho \sigma }\sum_{\beta }F_{(\beta )\mu
\rho }F_{(\beta )\kappa \sigma }^{\ast }.  \label{Cmu}
\end{equation}%
By taking into account that $C_{0}^{0}=-\langle E^{2}\rangle $ and using the
expression (\ref{L1}) for $\langle L\rangle $, for the topological part in
the VEV\ of the energy density one gets%
\begin{equation}
\langle T_{0}^{0}\rangle _{\mathrm{t}}=\frac{\Gamma (\left( D+1\right) /2)}{%
\left( 4\pi \right) ^{(D+1)/2}r^{D+1}}\left[ \left( D-3\right)
^{2}c_{D-1}(q)-\left( D-1\right) c_{D+1}(q)\right] .  \label{T00}
\end{equation}%
Depending on the parameters $q$ and $D$, the energy density (\ref{T00}) can
be either positive or negative.

For the components of (\ref{Cmu}) corresponding to the axial stresses one
gets (no summation over $l=3,\ldots ,D$)%
\begin{equation}
C_{l}^{l}=-\frac{8\pi }{D-3}\langle L\rangle +\frac{4(4\pi )^{1-D/2}q}{%
\Gamma \left( D/2\right) }\sideset{}{'}{\sum}_{m=0}^{\infty
}\int_{0}^{\infty }dk\,k^{D-1}\int_{0}^{\infty }d\gamma \frac{\gamma }{%
\omega }\left[ G_{qm}(\gamma r)+\frac{D-3}{2}J_{qm}^{2}(\gamma r)\right] .
\label{Clln}
\end{equation}%
The transformation of the second term in the right-hand side is done in a
way similar to that for the squared electric field and for the axial
stresses we find (no summation over $l$)%
\begin{equation}
\langle T_{l}^{l}\rangle _{\mathrm{t}}=\langle T_{0}^{0}\rangle _{\mathrm{t}%
},\;l=3,\ldots ,D.  \label{Tll}
\end{equation}%
This relation could also be directly obtained from the Lorentz invariance of
the problem with respect to the boosts along the directions $z^{l}$, $%
l=3,\ldots ,D$.

The off-diagonal components of the vacuum energy-momentum tensor vanish and
it remains to consider the VEVs of the radial and azimuthal stresses. For
the radial component of the tensor (\ref{Cmu}) one finds%
\begin{equation}
C_{1}^{1}=\frac{4(4\pi )^{1-D/2}q}{\Gamma (D/2-1)}\sideset{}{'}{\sum}%
_{m=0}^{\infty }\int dk\,k^{D-3}\int_{0}^{\infty }d\gamma \frac{\gamma ^{3}}{%
\omega }\left[ \left( D-1\right) J_{qm}^{\prime 2}(\gamma r)-G_{qm}(\gamma
r)+J_{qm}^{2}(\gamma r)\right] .  \label{C11}
\end{equation}%
The evaluation of the parts corresponding to the last two terms in the
square brackets in (\ref{C11}) is similar to that for the corresponding
parts on the squared electric field. For the remaining part, by using (\ref%
{IntRep}), we get%
\begin{equation}
\int_{0}^{\infty }dk\,k^{D-3}\int_{0}^{\infty }d\gamma \frac{\gamma ^{3}}{%
\omega }J_{qm}^{\prime 2}(\gamma r)=\frac{\Gamma (D/2-1)}{\sqrt{\pi }}%
\int_{0}^{\infty }\frac{dy}{y^{D-2}}\int_{0}^{\infty }d\gamma \gamma
^{3}\,e^{-\gamma ^{2}y^{2}}J_{qm}^{\prime 2}(\gamma r).  \label{Int5}
\end{equation}%
The integral is evaluated by using the relation
\begin{equation}
\int_{0}^{\infty }d\gamma \gamma ^{3}e^{-\gamma ^{2}y^{2}}J_{qm}^{\prime
2}(\gamma r)=\lim_{r\rightarrow r^{\prime }}\partial _{r}\partial
_{r^{\prime }}\int_{0}^{\infty }d\gamma \gamma e^{-\gamma
^{2}y^{2}}J_{qm}(\gamma r)J_{qm}(\gamma r^{\prime }),  \label{Int6}
\end{equation}%
and (\ref{Int1}). The further steps are similar to that for the VEVs of the
squared fields and are based on (\ref{Summ}). In this way one finds%
\begin{equation}
\langle T_{1}^{1}\rangle _{\mathrm{t}}=-\frac{\Gamma \left( (D+1)/2\right) }{%
\left( 4\pi \right) ^{(D+1)/2}r^{D+1}}\left( D-1\right) c_{D+1}(q).
\label{T11}
\end{equation}

Finally, for the component $C_{2}^{2}$ one has
\begin{equation}
C_{2}^{2}=\frac{4(4\pi )^{D/2-1}q}{\Gamma (D/2-1)}\sideset{}{'}{\sum}%
_{m=0}^{\infty }\int d\mathbf{k}\int_{0}^{\infty }d\gamma \frac{\gamma ^{3}}{%
\omega }\left[ \left( 1-D\right) J_{qm}^{\prime 2}(x)+\left( D-2\right)
G_{qm}(\gamma r)+J_{qm}^{2}(x)\right] ,  \label{C22}
\end{equation}%
and the evaluation is similar to the for (\ref{C11}). For the azimuthal
stress one gets%
\begin{equation}
\langle T_{2}^{2}\rangle _{\mathrm{t}}=\frac{\Gamma \left( (D+1)/2\right) }{%
\left( 4\pi \right) ^{(D+1)/2}r^{D+1}}D\left( D-1\right) c_{D+1}(q).
\label{T22}
\end{equation}%
As seen, one has the relation $\langle T_{2}^{2}\rangle _{\mathrm{t}%
}=-D\langle T_{1}^{1}\rangle _{\mathrm{t}}$. The VEV of the energy-momentum
tensor obeys the covariant conservation equation $\nabla _{\nu }\langle
T_{\mu }^{\nu }\rangle _{\mathrm{t}}=0$. For the geometry under
consideration the latter is reduced to the single equation $\partial
_{r}\left( r\langle T_{1}^{1}\rangle _{\mathrm{t}}\right) =\langle
T_{2}^{2}\rangle _{\mathrm{t}}$. For $D=3$ the vacuum energy-momentum tensor
is traceless, $\langle T_{\mu }^{\mu }\rangle _{\mathrm{t}}=0$. For $D\neq 3$
the electromagnetic field is not conformally invariant and the trace is not
zero.

For $D=3$, by taking into account (\ref{c6q}) one finds
\begin{equation}
\langle T_{\mu }^{\nu }\rangle _{\mathrm{t}}=-\frac{q^{2}-1}{720\pi ^{2}r^{4}%
}\left( q^{2}+11\right) \mathrm{diag}\left( 1,1,-3,1\right) .  \label{TD3}
\end{equation}%
In particular, the energy density is negative for $q>1$. This result was
obtained in \cite{Frol87,Dowk87} by using the corresponding Green function.
In the special case $D=5$ we have (no summation over $l$)
\begin{equation}
\langle T_{l}^{l}\rangle _{\mathrm{t}}=-\frac{\left( q^{2}-1\right) \left(
q^{2}+5\right) \left( q^{2}-4\right) }{945\left( 2\pi \right) ^{3}r^{6}},
\label{TllD5}
\end{equation}%
for $l=0,3,\ldots ,D$, and%
\begin{equation}
\langle T_{1}^{1}\rangle _{\mathrm{t}}=-\frac{1}{5}\langle T_{2}^{2}\rangle
_{\mathrm{t}}=-\frac{q^{2}-1}{1890\left( 2\pi \right) ^{3}r^{6}}%
(2q^{4}+23q^{2}+191).  \label{T11D5}
\end{equation}%
The energy density is positive for $1<q<2$ and negative for $q>2$. In the
special case $q=2$ and for general $D$ one obtains%
\begin{equation}
\langle T_{\mu }^{\nu }\rangle _{\mathrm{t}}=\frac{\left( D-2\right) \Gamma
(\left( D+1\right) /2)}{2\left( 4\pi \right) ^{(D+1)/2}r^{D+1}}\mathrm{diag}%
\left( D-5,-\frac{D-1}{D-2},D\frac{D-1}{D-2},D-5,\ldots ,D-5\right) .
\label{Tq2}
\end{equation}%
The corresponding energy density vanishes for $D=5$.

\begin{figure}[tbph]
\begin{center}
\epsfig{figure=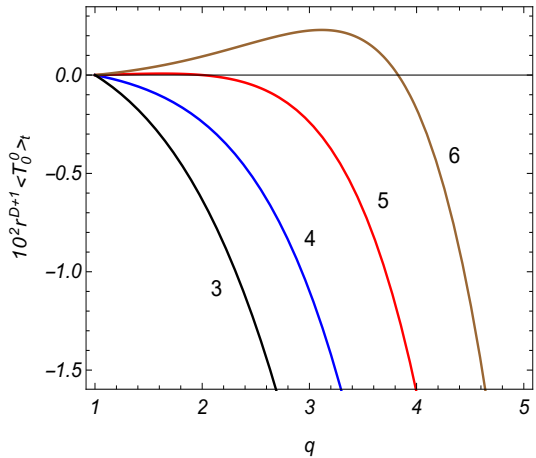,width=7.cm,height=6.cm}
\end{center}
\caption{The dependence of the vacuum energy density, multiplied by $%
10^2r^{D+1}$, on the parameter $q$ for different values of $D$ (the numbers
near the curves).}
\label{fig3}
\end{figure}

In figure \ref{fig3} we have plotted the VEV of the energy density,
multiplied by $r^{D+1}$, versus the parameter $q$ for different values of
the spatial dimension $D$ (the numbers near the curves). For $D\geqslant 5$
the energy density is positive for small values of $q$ and is negative for
large values of that parameter. For some intermediate value of $q$ there is a
maximum with positive energy density.

\section{VEVs for cosmic string on locally dS bulk}

\label{sec:dS}

In the second part of the paper we will be concerned about the combined
effects of a cosmic string and of the background gravitational field,
generated by a positive cosmological constant, on the VEVs of the squared
electric and magnetic fields and on the vacuum energy density. This
investigation has been recently presented in \cite{Saha17} based on the
corresponding two-point functions for the electromagnetic field tensor. Here
we follow a simpler approach based on the direct evaluation of the mode sums
introducing a cutoff function. The vacuum polarization induced by a cosmic
string in locally (but not globally) dS spacetime for massive scalar and
fermionic fields has been investigated in \cite{Beze09,Beze10,Moha15}.

Our choice of the locally dS geometry is motivated by several reasons. First
of all the dS spacetime is maximally symmetric and similar to case of the
flat background the problem for evaluation of the local characteristics of
the electromagnetic field is exactly solvable. In addition, the dS spacetime
plays an important role in modern cosmology. In most inflationary models the
early expansion of the universe is quasi de Sitterian (for the effects of
inflation on the cosmic strings see \cite{Basu94}-\cite{Avel07}). The
corresponding accelerated expansion before the radiation dominated era
naturally solves a number of problems in the standard cosmological model.
The inflation also provides an attractive mechanism of producing
long-wavelength electromagnetic fluctuations, originating from
subhorizon-sized quantum fluctuations of the electromagnetic field stretched
by the dS phase to superhorizon scales. In the post-inflationary era these
long-wavelength fluctuations re-enter the horizon and can serve as seeds for
cosmological magnetic fields. Related to this mechanism, the cosmological
dynamics of the electromagnetic field quantum fluctuations have been
discussed in large number of papers ( see \cite{Kron94}-\cite{Durr13} and
references therein). More recently, the observational data on high redshift
supernovae, galaxy clusters, and cosmic microwave background indicate that
at the present epoch the universe is accelerating and the corresponding
expansion is driven by a source the properties of which are close to a
positive cosmological constant. In this case, the quasi-dS geometry is the
future attractor for the universe. Though the cosmic strings produced before
or during early stages of the inflationary phase are diluted by the
quasi-exponential expansion, the defects can be formed near or at the end of
inflation by several mechanisms (see \cite{Hind11} for possible
observational consequences from this type of models). It is also possible to
have subsequent inflationary stages, with linear defects being formed in
between them \cite{Vile97}.

\subsection{Electromagnetic field modes}

In terms of the conformal time coordinate $\tau $, $-\infty <\tau <0$, the
line element describing the geometry under consideration is given by the
expression
\begin{equation}
ds^{2}=\left( \alpha /\tau \right) ^{2}[d\tau ^{2}-dr^{2}-r^{2}d\phi
^{2}-\left( d\mathbf{z}\right) ^{2}],  \label{ds2}
\end{equation}%
where the ranges for the spatial coordinates are the same as in (\ref{ds2M}%
). In the absence of cosmic string one has $\phi _{0}=2\pi $ and the
geometry coincides with the dS one in inflationary coordinates. For the
synchronous time $t$ one has $t=-\alpha \ln (\left\vert \tau \right\vert
/\alpha ),\ -\infty <t<+\infty $. The parameter $\alpha $ is related to the
cosmological constant $\Lambda $ by $\Lambda =D(D-1)/(2\alpha ^{2})$. It has
been argued in \cite{Ghez02,Abba03} that the vortex solution of the
Einstein-Abelian-Higgs equations in the presence of a cosmological constant
induces a deficit angle into dS spacetime. Similar to the case of the dS
spacetime, we can also write the line element (\ref{ds2}) with an angular
deficit in static coordinates. For simplicity considering the case $D=4$,
the corresponding transformation reads
\begin{equation}
t=t_{s}-\alpha \ln f(r_{s}),\;r=r_{s}f(r_{s})e^{-t_{s}/\alpha }\sin \theta
,\;z_{1}=r_{s}f(r_{s})e^{-t_{s}/\alpha }\cos \theta ,\;\phi =\phi ,
\label{Coord}
\end{equation}%
where $f(r_{s})=1/\sqrt{1-r_{s}^{2}/\alpha ^{2}}$. The line element takes
the form%
\begin{equation}
ds^{2}=f^{-2}(r_{s})dt_{s}^{2}-f^{2}(r_{s})dr_{s}^{2}-r_{s}^{2}(d\theta
^{2}+\sin ^{2}\theta d\phi ^{2}),  \label{dSstatic}
\end{equation}%
with $0\leq \phi \leq \phi _{0}$. With a new coordinate $\varphi =q\phi $,
from (\ref{dSstatic}) we obtain the static line element of dS
spacetime with deficit angle previously discussed in \cite{Ghez02}. In this
paper it is shown that to leading order in the gravitational coupling the
effect of the vortex on dS spacetime is to create a deficit angle in
the metric (\ref{dSstatic}).

For the coordinates corresponding to (\ref{ds2}) and for the Bunch-Davies
vacuum state the cylindrical electromagnetic modes are presented as%
\begin{eqnarray}
A_{(\beta )\mu }(x) &=&c_{\beta }\eta ^{\frac{D}{2}-1}H_{\frac{D}{2}%
-1}^{(1)}(\omega \eta )\left( 0,\frac{iqm}{r},-r\partial _{r},0,\ldots
,0\right) J_{q|m|}(\gamma r)e^{iqm\phi +i\mathbf{k}\cdot \mathbf{z}%
},\;\sigma =1,  \notag \\
A_{(\beta )\mu }(x) &=&c_{\beta }\omega \eta ^{\frac{D}{2}-1}H_{\frac{D}{2}%
-1}^{(1)}(\omega \eta )\left( 0,\epsilon _{\sigma l}+i\frac{\mathbf{k}\cdot
\boldsymbol{\epsilon }_{\sigma }}{\omega ^{2}}\partial _{l}\right)
J_{q|m|}(\gamma r)e^{iqm\phi +i\mathbf{k}\cdot \mathbf{z}},\;\sigma
=2,\ldots ,D-1,  \label{AmudS}
\end{eqnarray}%
where $H_{\nu }^{(1)}(x)$ is the Hankel function of the first kind and $\eta
=|\tau |$. Other notations in (\ref{AmudS}) are the same as in (\ref{A1M}).
For the normalization constant from (\ref{NormCond}) one finds%
\begin{equation}
|c_{\beta }|^{2}=\frac{q}{4(2\pi \alpha )^{D-3}\gamma }.  \label{cbetdS}
\end{equation}%
For $D=3$ the electromagnetic field is conformally invariant and the modes (%
\ref{AmudS}) coincide with (\ref{A1M}).

\subsection{Squared electric field}

We start with the VEV of the squared electric field obtained from (\ref%
{E2mode}). The VEV, regularized by introducing the cutoff function $%
e^{-b\omega ^{2}}$, is presented as%
\begin{eqnarray}
\langle E^{2}\rangle _{\mathrm{reg}} &=&\frac{2^{5-D}q\eta ^{D+2}}{\pi
^{D/2}\Gamma (D/2-1)\alpha ^{D+1}}\sideset{}{'}{\sum}_{m=0}^{\infty
}\int_{0}^{\infty }dk\,k^{D-3}\int_{0}^{\infty }d\gamma \,\gamma |K_{\nu
}(e^{i\pi /2}\omega \eta )|^{2}  \notag \\
&&\times e^{-b\omega ^{2}}\left\{ \left[ (D-2)\gamma ^{2}+(D-3)k^{2}\right]
J_{qm}^{2}(\gamma r)+\left( \gamma ^{2}+2k^{2}\right) G_{qm}(\gamma
r)\right\} ,  \label{E2S}
\end{eqnarray}%
where we have introduced the MacDonald function $K_{\nu }(x)$ instead of the
Hankel function, $\nu =D/2-2$ and the other notations are the same as in (%
\ref{E2c}). For the further transformation we use the integral representation%
\begin{equation}
|K_{\nu }(e^{i\pi /2}\eta \omega )|^{2}=\frac{1}{2}\int_{0}^{\infty
}dy\,\cosh (\nu y)\int_{0}^{\infty }\frac{dx}{x}e^{-2\eta ^{2}\left( \cosh
y-1\right) /x-x\omega ^{2}/4}.  \label{Kint}
\end{equation}%
This relation is obtained from the integral representation of the product of
MacDonald functions with different arguments given in \cite{Wats66}.
Plugging (\ref{Kint}) into (\ref{E2S}), the integration over $k$ is
elementary and the integral over $y$ is expressed in terms of the function $%
K_{\nu }(2\eta ^{2}/x)$. In this way one gets%
\begin{eqnarray}
\langle E^{2}\rangle _{\mathrm{reg}} &=&\frac{8q\eta ^{D+2}}{\left( 4\pi
\right) ^{D/2}\alpha ^{D+1}}\sideset{}{'}{\sum}_{m=0}^{\infty
}\int_{0}^{\infty }dx\frac{e^{2\eta ^{2}/x}}{xw^{D/2}}K_{\nu }(2\eta
^{2}/x)\int_{0}^{\infty }d\gamma \gamma  \notag \\
&&\times \left[ \left( D-2+w\partial _{w}\right) G_{qm}(\gamma
r)+(D-2)\left( \frac{D-3}{2}-w\partial _{w}\right) J_{qm}^{2}(\gamma r)%
\right] e^{-w\gamma ^{2}},  \label{E2S1}
\end{eqnarray}%
where $w=x/4+b$. The integrals over $\gamma $ are evaluated by using the
formulas (\ref{Int1}), (\ref{Int4}) and we find%
\begin{eqnarray}
\langle E^{2}\rangle _{\mathrm{reg}} &=&\frac{8q(\eta /r)^{D+2}}{\left( 2\pi
\right) ^{D/2}\alpha ^{D+1}}\int_{0}^{\infty }\frac{dx}{x}e^{2\eta
^{2}/x}K_{D/2-2}(2\eta ^{2}/x)u^{D/2+1}  \notag \\
&&\times \left[ \left( D-2+\partial _{u}u\right) \left( \partial
_{u}+1\right) +(D-2)\left( \frac{D-3}{2}+\partial _{u}u\right) \right] e^{-u}%
\sideset{}{'}{\sum}_{m=0}^{\infty }I_{qm}(u),  \label{E2S2}
\end{eqnarray}%
with $u=r^{2}/[2\left( x/4+b\right) ]$.

Next we use the formula (\ref{Summ}) for the series over $m$. The $l=0$ term
gives the regularized VEV\ in dS spacetime in the absence of the cosmic
string, denoted here as $\langle E^{2}\rangle _{\mathrm{dS}}^{\mathrm{reg}}$%
. The remaining part corresponds to the contribution of the cosmic string.
For points $r\neq 0$ that part is finite in the limit $b\rightarrow 0$ and
the cutoff can be removed. In this way, for the topological part $\langle
E^{2}\rangle _{\mathrm{t}}=\langle E^{2}\rangle -\langle E^{2}\rangle _{%
\mathrm{dS}}$ one gets \cite{Saha17}%
\begin{equation}
\langle E^{2}\rangle _{\mathrm{t}}=\frac{8\alpha ^{-D-1}}{(2\pi )^{D/2}}%
\left[ \sideset{}{'}{\sum}_{l=1}^{[q/2]}g_{E}(r/\eta ,s_{l})-\frac{q}{\pi }%
\sin (q\pi )\int_{0}^{\infty }dy\frac{g_{E}(r/\eta ,\cosh y)}{\cosh
(2qy)-\cos (q\pi )}\right] ,  \label{E2S3}
\end{equation}%
where $\langle E^{2}\rangle _{\mathrm{dS}}$ is the renormalized VEV in the
absence of the cosmic string. In (\ref{E2S3}) we have introduced the
notation $s_{l}=\sin (\pi l/q)$ and
\begin{equation}
g_{E}(x,y)=\int_{0}^{\infty }du\,u^{D/2}K_{D/2-2}(u)e^{u-2x^{2}y^{2}u}\left[
2ux^{2}y^{2}\left( 2y^{2}-D+1\right) +\left( D-1\right) \left(
D/2-2y^{2}\right) \right] .  \label{gE}
\end{equation}%
Note that the regularized VEV for dS bulk in the absence of cosmic string
has the form%
\begin{equation}
\langle E^{2}\rangle _{\mathrm{dS}}^{\mathrm{reg}}=\frac{2D\left( D-1\right)
}{\left( 2\pi \right) ^{D/2}\alpha ^{D+1}}\int_{0}^{\infty }dz\frac{%
z^{D/2}e^{z}K_{D/2-2}(z)}{\left( 1+2zb/\eta ^{2}\right) ^{D/2+1}}.
\label{E2reg}
\end{equation}%
As expected, this VEV\ diverges in the limit $b\rightarrow 0$ and requires a
renormalization. From the maximal symmetry of dS spacetime and of the
Bunch-Davies vacuum state we expect that the renormalized VEV does not
depend on the spacetime point and $\langle E^{2}\rangle _{\mathrm{dS}}=%
\mathrm{const}\cdot \alpha ^{-D-1}$. In (\ref{E2S3}), the topological part
depends on the coordinates $r$ and $\eta $ in the form of the ratio $r/\eta $%
. The latter is the proper distance from the string, $\alpha r/\eta $,
measured in units of the dS curvature scale $\alpha $.

For odd values of $D$ the function $g_{E}(x,y)$ in (\ref{E2S3}) is expressed
in terms of the elementary functions. In particular, one finds%
\begin{eqnarray}
\langle E^{2}\rangle _{\mathrm{t}} &=&-\frac{\left( q^{2}-1\right) \left(
q^{2}+11\right) }{180\pi (\alpha r/\eta )^{4}},\;D=3,  \notag \\
\langle E^{2}\rangle _{\mathrm{t}} &=&-\frac{\left( q^{2}-1\right) \left(
q^{4}+22q^{2}+211\right) }{1890\pi ^{2}\left( \alpha r/\eta \right) ^{6}}%
,\;D=5.  \label{E2SD5}
\end{eqnarray}%
For $D=3$ the electromagnetic field is conformally invariant and the
topological contribution is related to the one for the flat bulk by
standard conformal relation. It is of interest to note that a similar
relation takes place for $D=5$ as well. For other values of $D$ there is no
such a simple relation.

For points near the string, assuming that $r/\eta \ll 1$, the contribution
of the large $u$ is dominant in (\ref{gE}). By using the corresponding
asymptotic expression for the MacDonald function, to the leading order we
get $\langle E^{2}\rangle _{\mathrm{t}}\approx (\eta /\alpha )^{D+1}\langle
E^{2}\rangle _{\mathrm{t}}^{\mathrm{(M)}}$, where $\langle E^{2}\rangle _{%
\mathrm{t}}^{\mathrm{(M)}}$ is topological contribution in the flat
bulk and is given by (\ref{E23}). Near the string the main contribution to
the VEVs comes from the fluctuations with wavelengths smaller than the
curvature radius and the influence of the background gravitational field on
the corresponding modes is weak.

At proper distances from the string larger than the dS curvature radius one
has $r/\eta \gg 1$. Now, the contribution from the region near the lower
limit of the integration in (\ref{gE}) is dominant. For $D\geqslant 5$, to
the leading order, the topological part is expressed in terms of $c_{4}(q)$
and $c_{6}(q)$:%
\begin{equation}
\langle E^{2}\rangle _{\mathrm{t}}\approx \frac{\Gamma \left( D/2-2\right)
\left( q^{2}-1\right) }{8\pi ^{D/2}\alpha ^{D+1}\left( r/\eta \right) ^{6}}%
\left[ \frac{D-1}{21}\left( D-6\right) (2q^{4}+23q^{2}+191)-4\left(
D-4\right) \left( q^{2}+11\right) \right] .  \label{E2far}
\end{equation}%
In the special case $D=4$ from (\ref{E2S3}) we get%
\begin{equation}
\langle E^{2}\rangle _{\mathrm{t}}\approx -\frac{\left( q^{2}-1\right) \ln
(r/\eta )}{630\pi ^{2}\alpha ^{5}(r/\eta )^{6}}(2q^{4}+23q^{2}+191).
\label{E2far2}
\end{equation}%
For $D=3$ one has the behavior given by (\ref{E2SD5}). At large distances,
the total VEV is dominated by the part $\langle E^{2}\rangle _{\mathrm{dS}}$%
. As seen, in spatial dimensions $D=4,6,7,\ldots $, at distances larger than
the curvature radius, the influence of the gravitational field on the
topological part is essential. From (\ref{E2far}) and (\ref{E2far2}) it
follows that at large distances $\langle E^{2}\rangle _{\mathrm{t}}<0$ for $%
D=4,5,6$, and $\langle E^{2}\rangle _{\mathrm{t}}>0$ for $D\geqslant 7$. By
taking into account that near the string one has $\langle E^{2}\rangle _{%
\mathrm{t}}<0$ for $D\geqslant 3$ we conclude that in spatial dimensions $%
D\geqslant 7$ the topological contribution $\langle E^{2}\rangle _{\mathrm{t}%
}$ has a positive maximum for some intermediate value of $r/\eta $.

\subsection{Squared magnetic field}

By using the mode functions (\ref{AmudS}), from the mode-sum formula (\ref%
{B2}) one gets the following representation for the squared magnetic field%
\begin{eqnarray}
\langle B^{2}\rangle _{\mathrm{reg}} &=&\frac{2^{5-D}\pi ^{-D/2}q\eta ^{D+2}%
}{\Gamma (D/2-1)\alpha ^{D+1}}\sideset{}{'}{\sum}_{m=0}^{\infty
}\int_{0}^{\infty }dk\,k^{D-3}\int_{0}^{\infty }d\gamma \,\gamma
K_{D/2-1}(e^{-i\pi /2}\omega \eta )K_{D/2-1}(e^{i\pi /2}\omega \eta )  \notag
\\
&&\times e^{-b\omega ^{2}}\left\{ \left[ \left( D-2\right) \gamma ^{2}+2k^{2}%
\right] G_{qm}(\gamma r)+\left[ \gamma ^{2}+(D-3)k^{2}\right]
J_{qm}^{2}(\gamma r)\right\} .  \label{B2S}
\end{eqnarray}%
Further transformation of this VEV is similar to that for the electric field
squared. The final formula for the topological contribution $\langle
B^{2}\rangle _{\mathrm{t}}=\langle B^{2}\rangle -\langle B^{2}\rangle _{%
\mathrm{dS}}$ is given by \cite{Saha17}
\begin{equation}
\langle B^{2}\rangle _{\mathrm{t}}=\frac{8\alpha ^{-D-1}}{(2\pi )^{D/2}}%
\left[ \sideset{}{'}{\sum}_{l=1}^{[q/2]}g_{M}(r/\eta ,s_{l})-\frac{q}{\pi }%
\sin (q\pi )\int_{0}^{\infty }dy\frac{g_{M}(r/\eta ,\cosh y)}{\cosh
(2qy)-\cos (q\pi )}\right] ,  \label{B2S1}
\end{equation}%
with the function%
\begin{eqnarray}
g_{M}(x,y) &=&\int_{0}^{\infty
}du\,u^{D/2}K_{D/2-1}(u)e^{u-2x^{2}y^{2}u}\left\{ (D-1)D/2\right.  \notag \\
&&\left. -4(D-2)y^{2}+2x^{2}y^{2}u\left[ 2(D-2)y^{2}-D+1\right] \right\} .
\label{gMn}
\end{eqnarray}%
We denote by $\langle B^{2}\rangle _{\mathrm{dS}}$ the renormalized VEV of
the squared magnetic field in dS spacetime in the absence of the cosmic
string. The regularized VEV of the squared magnetic field for the latter
geometry is given by
\begin{equation}
\langle B^{2}\rangle _{\mathrm{dS}}^{\mathrm{reg}}=\frac{2D\left( D-1\right)
}{\left( 2\pi \right) ^{D/2}\alpha ^{D+1}}\int_{0}^{\infty }dz\frac{%
z^{D/2}e^{z}K_{D/2-1}(z)}{\left( 1+2zb/\eta ^{2}\right) ^{D/2+1}}.
\label{B2reg}
\end{equation}%
For $D=3$ this result coincides with that for the squared electric field.
Note that for $D=4$ and $q=3$ the topological part vanishes, $\langle
B^{2}\rangle _{\mathrm{t}}=0$.

For odd values of the spatial dimension $D$ the integral in (\ref{gMn}) is
expressed in terms of the elementary functions. For $D=3$ one has $\langle
B^{2}\rangle _{\mathrm{t}}=\langle E^{2}\rangle _{\mathrm{t}}$ and for $D=5$
we get%
\begin{equation}
\langle B^{2}\rangle _{\mathrm{t}}=\frac{\left( 3+r^{2}/\eta ^{2}\right)
c_{4}(q)-c_{6}(q)}{2\pi ^{2}\left( \alpha r/\eta \right) ^{6}}.  \label{B25}
\end{equation}%
Note that, unlike to the case of the electric field, the VEV\ of the squared
magnetic field for $D=5$ does not coincide with the corresponding VEV in the
flat bulk with the distance $r$ replaced by the proper distance $\alpha
r/\eta $.

For points near the string, $r/\eta \ll 1$, the influence of the
gravitational field on the topological contribution is weak and to the
leading order one has the relation $\langle B^{2}\rangle _{\mathrm{t}%
}\approx \left( \eta /\alpha \right) ^{D+1}\langle B^{2}\rangle _{\mathrm{t}%
}^{\mathrm{(M)}}$, where the VEV $\langle B^{2}\rangle _{\mathrm{%
t}}^{\mathrm{(M)}}$ is given by the expression (\ref{B22}). The influence of
the gravitational field is essential at proper distances from the string
larger than the curvature radius of dS spacetime. In the asymptotic region $%
r/\eta \gg 1$ for the leading terms in the expansions of the topological
parts one has%
\begin{eqnarray}
\langle B^{2}\rangle _{\mathrm{t}} &\approx &-\frac{\left( q^{2}-1\right)
\left( q^{2}-9\right) \left( 2q^{2}+13\right) }{1260\pi ^{2}\alpha
^{5}\left( r/\eta \right) ^{6}},\;D=4,  \notag \\
\langle B^{2}\rangle _{\mathrm{t}} &\approx &\frac{(D-1)\left( D-4\right)
\Gamma (D/2-1)}{4\pi ^{D/2}\alpha ^{D+1}(r/\eta )^{4}}c_{4}(q),\;D\neq 4.
\label{B2l}
\end{eqnarray}%
Note that for $D=4$ at large distances one has $\langle E^{2}\rangle _{%
\mathrm{t}}/\langle B^{2}\rangle _{\mathrm{t}}\propto \ln (r/\eta )$,
whereas for $D\geqslant 5$ one has $\langle E^{2}\rangle _{\mathrm{t}%
}/\langle B^{2}\rangle _{\mathrm{t}}\propto (r/\eta )^{-2}$ and the
topological part in the VEV\ of the squared magnetic field is much larger
than the one for the electric field. For $D\geqslant 5$ the topological part
$\langle B^{2}\rangle _{\mathrm{t}}$ is positive at large distances.

\begin{figure}[tbph]
\begin{center}
\epsfig{figure=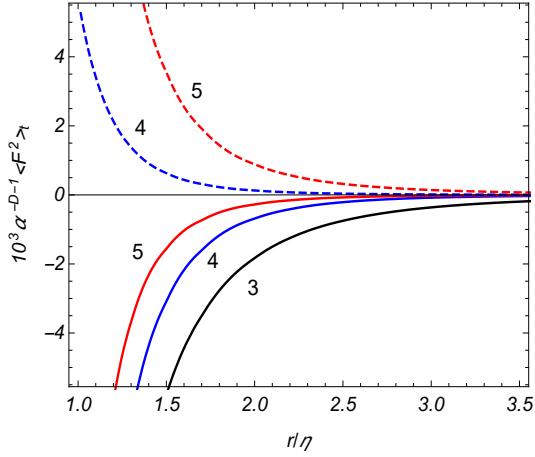,width=7.cm,height=6.cm}
\end{center}
\caption{The topological contributions in the VEVs of the squared electric
and magnetic fields on dS bulk versus $r/\protect\eta $ for different values
of $D$ (the numbers near the curves).}
\label{fig4}
\end{figure}

In figure \ref{fig4} we have displayed the dependence of the VEVs for
squared electric (full curves) and magnetic (dashed curves) fields on the
ratio $r/\eta $ (proper distance from the string in units of the dS
curvature scale $\alpha $) for separate values of the spatial dimension $%
D=3,4,5$. For $D=3$ the VEVs for the electric and magnetic fields coincide.
The graphs are plotted for $q=1.5$.

\subsection{VEV of the energy density}

The VEV of the energy density is given by $\langle T_{0}^{0}\rangle =\left(
\langle E^{2}\rangle +\langle B^{2}\rangle \right) /8\pi $ and is decomposed
as $\langle T_{0}^{0}\rangle =\langle T_{0}^{0}\rangle _{\mathrm{dS}%
}+\langle T_{0}^{0}\rangle _{\mathrm{t}}$, where $\langle T_{\mu }^{\nu
}\rangle _{\mathrm{dS}}=\mathrm{const}\cdot \delta _{\mu }^{\nu }$ is the
VEV in dS spacetime in the absence of the cosmic string. For the topological
contribution one gets the expression%
\begin{equation}
\langle T_{0}^{0}\rangle _{\mathrm{t}}=\frac{2\alpha ^{-D-1}}{(2\pi )^{D/2+1}%
}\left[ \sideset{}{'}{\sum}_{l=1}^{[q/2]}g(r/\eta ,s_{l})-\frac{q}{\pi }\sin
(q\pi )\int_{0}^{\infty }dy\frac{g(r/\eta ,\cosh y)}{\cosh (2qy)-\cos (q\pi )%
}\right] ,  \label{eps1}
\end{equation}%
with the function%
\begin{eqnarray}
g(x,y) &=&\int_{0}^{\infty }du\,u^{D/2}e^{u-2x^{2}y^{2}u}\left\{ K_{\frac{D}{%
2}-2}(u)\left[ \left( D-1\right) \left( \frac{D}{2}-2y^{2}\right)
+2x^{2}y^{2}u\left( 2y^{2}-D+1\right) \right] \right.  \notag \\
&&\left. +K_{\frac{D}{2}-1}(u)\left[ (D-1)\frac{D}{2}%
-4(D-2)y^{2}+2x^{2}y^{2}u\left( 2(D-2)y^{2}-D+1\right) \right] \right\} .
\label{gxy}
\end{eqnarray}%
Near the string, $r/\eta \ll 1$, to the leading order, one has $\langle
T_{0}^{0}\rangle _{\mathrm{t}}\approx (\eta /\alpha )^{D+1}\langle
T_{0}^{0}\rangle _{\mathrm{t}}^{\mathrm{(M)}}$ with the energy
density in the flat spacetime from (\ref{T00}). At large distances from the string one has the
asymptotic expressions%
\begin{eqnarray}
\langle T_{0}^{0}\rangle _{\mathrm{t}} &\approx &-\frac{\left(
q^{2}-1\right) \ln (r/\eta )}{5040\pi ^{3}\alpha ^{5}(r/\eta )^{6}}%
(2q^{4}+23q^{2}+191),\;D=4,  \notag \\
\langle T_{0}^{0}\rangle _{\mathrm{t}} &\approx &\frac{(D-1)\left(
D-4\right) \Gamma (D/2-1)}{2880\pi ^{D/2+1}\alpha ^{D+1}(r/\eta )^{4}}\left(
q^{2}-1\right) \left( q^{2}+11\right) ,\;D\geqslant 5.  \label{T00L}
\end{eqnarray}%
In this asymptotic region the vacuum energy density is dominated by the part
$\langle T_{0}^{0}\rangle _{\mathrm{dS}}$. The topological contribution is
negative for $D=4$ and positive for $D\geqslant 5$.

\begin{figure}[tbph]
\begin{center}
\epsfig{figure=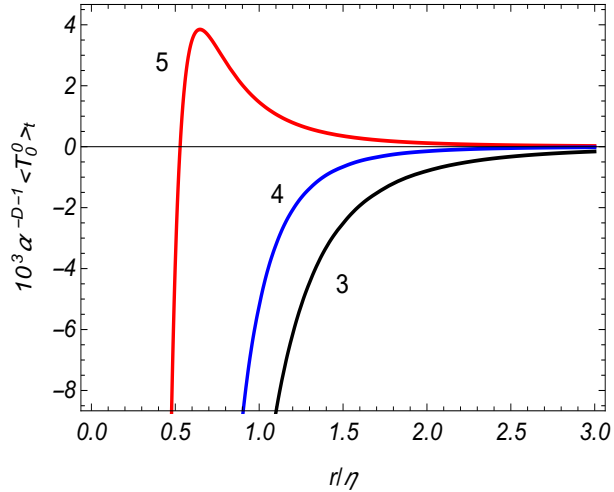,width=8.cm,height=6.5cm}
\end{center}
\caption{The topological part in the vacuum energy density on dS bulk as a
function of $r/\protect\eta $, for different values of $D$ (the numbers near
the curves). The graphs are plotted for $q=2.5$.}
\label{fig5}
\end{figure}

Figure \ref{fig5} presents the energy density versus $r/\eta $ for different
values of the spatial dimension $D=3,4,5$.

One of the interesting effects during the dS expansion, playing an important
role in inflationary cosmology, is the so-called classicalization of quantum
fluctuations: the evolution of quantum fluctuations into classical
fluctuations. In particular, the latter for the inflaton field are expected
to be the seeds of large scale structures in the universe. In a similar way,
the classicalization of the electromagnetic fluctuations may give rise to
large scale magnetic fields (for various types of mechanisms for the
generation of cosmological magnetic fields see, for instance, \cite{Durr13,Kron94,Giov04,Kand11}). 
The discussion we have presented above shows that these
fields will be influenced by cosmic strings formed at late stages of the
inflationary phase.

\section{Conclusion}

\label{sec:Conc}

We have investigated the influence of a straight cosmic string on the local
characteristics of the electromagnetic vacuum. First we have considered a
cosmic string on flat spacetime and then the corresponding results are
generalized for the locally dS background geometry. A simplified model is used where
the effect of the cosmic string on the background geometry is reduced to the
generation of a planar angle deficit. In this model, for points outside the
core the local geometry is not changed by the cosmic string, but the global
properties are different. The corresponding nontrivial topology gives rise
to shifts in the VEVs of physical observables. For the investigation of the
topological contributions we have employed the direct summation over the
complete set of electromagnetic modes.

For a cosmic string on $(D+1)$-dimensional flat spacetime the complete
set of modes for the vector potential is given by (\ref{A1M}), where $\sigma
$ enumerates the polarization states. For the regularization of the mode
sums in the VEVs of squared electric and magnetic fields the cutoff function
$e^{-b\omega ^{2}}$ is introduced. The application of the formula (\ref{Summ}%
) for the summation over the azimuthal quantum number allowed us to extract
explicitly the parts in the VEVs corresponding to the Minkwoski spacetime in
the absence of the cosmic string. For points away from the string core the
remaining topological contributions are finite in the limit $b\rightarrow 0$
and the cutoff can be safely removed. These contributions for the electric
and magnetic fields are given by the expressions (\ref{E23}) and (\ref{B22}%
), where the function $c_{n}(q)$ depending on the planar angle deficit is
defined by (\ref{cnqu}). In spatial dimensions $D\geqslant 3$ the
topological part in the VEV of the squared electric field is negative
whereas, depending on $q$ and $D$, the corresponding part for the magnetic
field can be either negative or positive. For odd values of $D$, the
functions $c_{n}(q)$ in the expressions for the topological contributions
are polynomials in $q$ and the VEVs are further simplified. In the special
case $D=3$ the electromagnetic field is conformally invariant and the
topological contributions for the electric and magnetic fields coincide.

Another important characteristic of the vacuum state that determines the
back-reaction of quantum effects on the background geometry is the VEV of
the energy-momentum tensor. For a cosmic string on flat bulk this VEV
is diagonal. Due to the Lorentz invariance with respect to the boosts along
directions $z^{l}$, $l=3,\ldots ,D$, the topological contributions in the
corresponding stresses coincide with the energy density, given by (\ref{T00}%
). The radial and azimuthal stresses are given by the expressions (\ref{T11}%
) and (\ref{T22}). They are related by a simple relation $\langle
T_{2}^{2}\rangle _{\mathrm{t}}=-D\langle T_{1}^{1}\rangle _{\mathrm{t}}$
that is a direct consequence of the covariant conservation equation for the
VEV\ of the energy-momentum tensor. The general expressions are simplified
for an odd number od spatial dimension. In particular, for $D=3,5$ one gets
the representations (\ref{TD3})-(\ref{T11D5}). Depending on the planar angle
deficit and the spatial dimension, the corresponding energy density can be
either negative or positive.

For a string in locally dS spacetime the electromagnetic mode functions, realizing
the Bunch-Davies vacuum state, are presented as (\ref{AmudS}). The
topological contributions in the VEVs of the squared electric and magnetic
fields are given by the formulas (\ref{E2S3}) and (\ref{B2S1}) with the
functions (\ref{gE}), (\ref{gMn}). The corresponding energy density is
obtained by summing the contributions of the electric and magnetic parts.
For points near the string the dominant contribution to the VEVs come from
the fluctuations with small wavelengths and the influence of the
gravitational field is weak. In this region, the leading terms in the VEVs
coincide with those for flat spacetime with the distance from the string
replaced by the proper distance. The influence of the gravitational field is
essential at proper distances larger than the curvature radius of dS
spacetime. For $D=4$ the topological contributions in the VEVs decay as $%
(r/\eta )^{-6}\ln (r/\eta )$ for the squared electric field and like $%
(r/\eta )^{-6}$ for the squared magnetic field. In this case the
corresponding energy density is dominated by the electrical part and is
negative. For $D>4$ the topological contributions decay as $(r/\eta )^{-6}$
in the case of the electric field squared and as $(r/\eta )^{-4}$ for the
magnetic field squared. The topological term in the vacuum energy density is
dominated by the magnetic part and is positive. This behavior is in clear
contrast to the problem on flat spacetime where the energy density decays
like $r^{-D-1}$.

\end{document}